\documentclass[english,twocolumn,prl,amsmath,amssymb,floatfix]{revtex4-1}
\pdfpageattr {/Group << /S /Transparency /I true /CS /DeviceRGB>>}
\usepackage[T1]{fontenc}
\usepackage[latin9]{inputenc}
\usepackage{dcolumn}
\usepackage{bm}
\usepackage{graphicx}
\usepackage{color}
\usepackage{babel}
\usepackage{amsfonts}
\usepackage{slashed}
\usepackage{enumerate}
\usepackage{braket}

\begin{document}


\title{Machine learning effective models for quantum systems}

\author{Jonas B. Rigo}
\author{Andrew K. Mitchell}
\affiliation{School of Physics, University College Dublin, Belfield, Dublin 4, Ireland}


\begin{abstract}
\noindent The construction of good effective models is an essential part of understanding and simulating complex systems in many areas of science. It is a particular challenge for correlated many body quantum systems displaying emergent physics. 
We propose a machine learning approach that optimizes an effective model based on an estimation of its partition function. 
The success of the method is demonstrated by application to the single impurity Anderson model and double quantum dots, where non-perturbative results are obtained for the old problem of mapping to effective Kondo models. 
We also show that an alternative approach based on learning minimal models from observables may yield the wrong low-energy physics. On the other hand, learning minimal models from the partition function recovers the correct low-energy physics but may not reproduce all observables.
\end{abstract}
\maketitle


The approach to understanding and simulating complex quantum systems can be divided into two groups: \textit{ab initio} studies in which one tries to account for all microscopic details, or studies of simplified effective models that still capture the essential physical phenomena of interest. A prerequisite for the latter is to construct a good effective model. The question of how to do this systematically, starting from a more complex microscopic system, is an important one for many areas of physics. 

Effective models are often defined in a reduced Hilbert space involving only those degrees of freedom relevant to describe the low-temperature physics of a complex microscopic model. They can be derived by perturbatively eliminating degrees of freedom, coarse-graining, or by using renormalization group (RG) methods \cite{Cardy1996xt,PhysRevB.4.3174,PhysRev.149.491,RevModPhys.47.773,*RevModPhys.80.395}: at low energies, microscopic details only enter through effective interactions and renormalized coupling constants. For the purposes of simulation, and to make realistic contact with experiment, the parameters as well as the structure of an effective model must be determined.

\begin{figure}[b]
  \centering
\includegraphics[width=8.5cm]{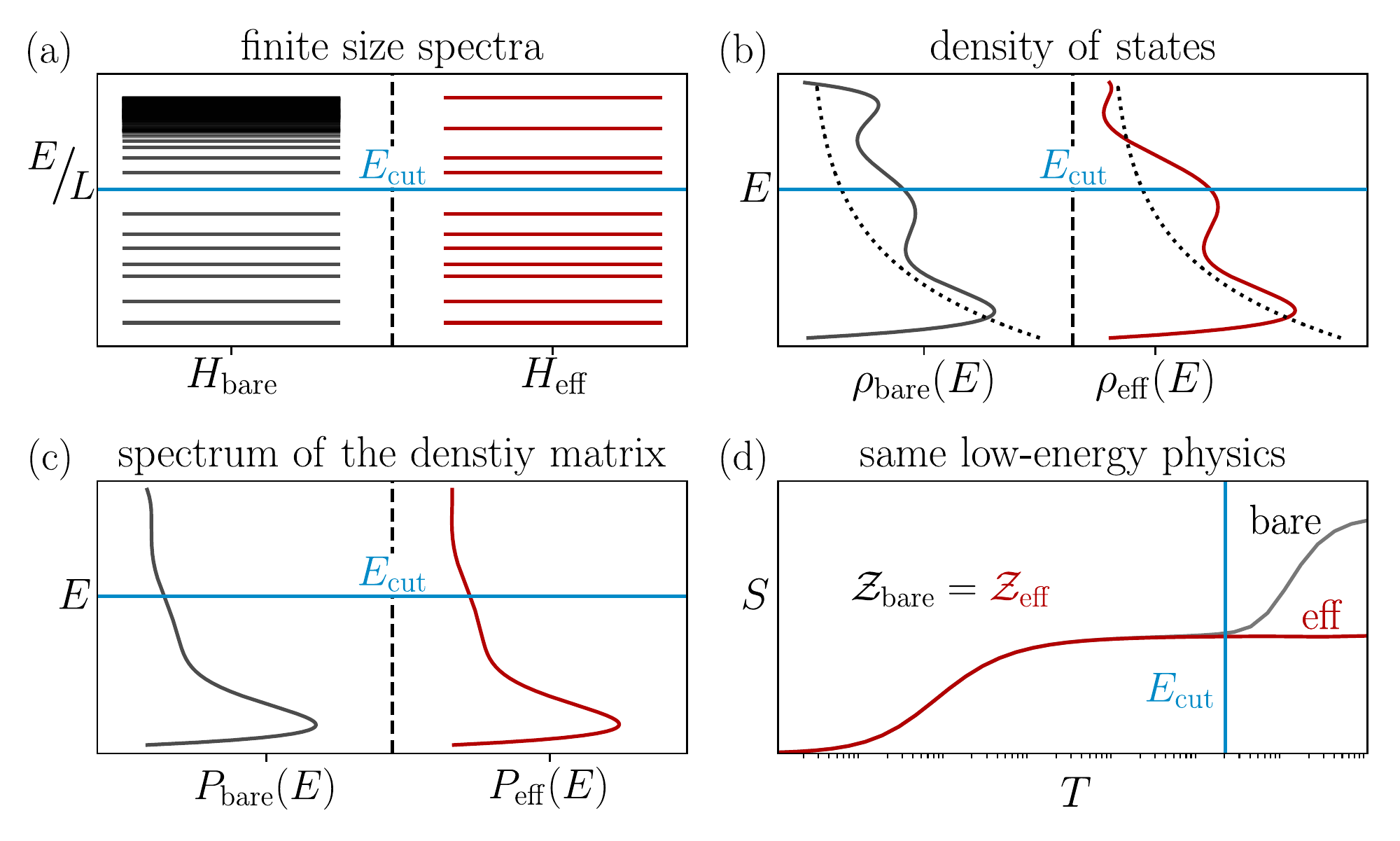}
  \caption{Schematic comparison of bare [left] and effective [right] models: (a) Finite size spectra, which agree up to some high-energy cutoff, $E_{\text{cut}}$. (b) Density of states (Boltzmann weighting $e^{-\beta E}$ as dotted line). (c) Spectrum of the thermal density matrix. (d) Thermodynamics, such as entropy, match at low temperatures $T\ll E_{\text{cut}}$ when $\mathcal{Z}_{\rm{eff}}=\mathcal{Z}_{\rm{bare}}$. }
  \label{fig:schematic}
\end{figure}

In this Letter, we use information theory and machine learning (ML) methods to find good effective models for quantum many-body systems. Our `model ML' approach is based on comparing the low-energy eigenspectrum of the effective and microscopic models (see Fig.~\ref{fig:schematic}), which gives a simple optimization condition on their partition functions. We compare this to an alternative approach involving learning from local observables. 
The two methods only agree at the Gibbs-Bogoliubov-Feynman \cite{feynman1998statistical} (GBF) bound. For \emph{minimal} effective models comprising only RG-relevant terms, learning from observables may not yield the same effective model parameters compared with learning from the partition function, since observables can flow under RG while the partition function does not \cite{Cardy1996xt,PhysRevB.4.3174,PhysRevA.92.022330}. Although the correct low-energy physics is obtained when the partition functions of bare and effective models agree, local observables may differ. However, minimally-constrained effective models, including marginal and irrelevant RG corrections, may reproduce the correct low-energy physics as well as the correct value of observables. Applications of ML optimization using observables have been discussed in Refs.~\cite{1907.05898,PhysRevB.95.041101,PhysRevB.95.241104}.

Effective models are also used in self-learning Monte Carlo \cite{PhysRevB.96.161102,PhysRevB.95.041101,PhysRevB.95.241104,PhysRevB.97.205140}. ML is used to optimize effective Hamiltonians that can be treated more efficiently while reproducing the same Monte Carlo update weights as the bare Hamiltonian for sampled configurations.
Such methods are not straightforwardly applicable to systems where the effective model involves rather different degrees of freedom, or the emergent physics is non-perturbative.

The full potential for applying ML concepts in physics is still being explored. Intense recent activity in the field covers a diverse range of topics, including: 
(i) finding and describing eigenstates  \cite{carleo,PhysRevLett.122.250501,PhysRevB.97.035116,1904.00031,PhysRevX.8.011006,1807.03325,1905.04312}; (ii) the inverse problem of finding parent Hamiltonians  \cite{PhysRevLett.112.190501,PhysRevResearch.1.033092,PhysRevLett.122.020504,PhysRevB.97.075114}; (iii) predicting properties of materials \cite{Butler2018,1906.08534,1907.03055}; (iv) identifying phases of matter \cite{Carrasquilla2017,PRL.121.245701,Zhang2019,PhysRevX.7.031038,1907.05417,PhysRevB.94.195105,Rem2019}; (v) improving numerical simulations \cite{PhysRevB.90.155136,PhysRevB.96.161102,PhysRevB.95.041101,PhysRevB.95.241104,PhysRevB.97.205140,PhysRevB.95.035105,PhysRevE.95.031301,PhysRevB.97.085104,PhysRevLett.108.253002,PhysRevLett.108.058301,PhysRevB.100.045153}; (vi) connections between ML and RG \cite{1906.05212,1410.3831,PhysRevLett.121.260601,Koch-Janusz2018,B_ny_2015,APENKO201262}.

Concepts from physics are also often used in ML algorithms \cite{Biamonte2017,PhysRevLett.117.130501,PhysRevX.8.021050} -- perhaps most notably in Boltzmann machines \cite{bengio,Bishop2006} where an unknown probability distribution is approximated by the physical Boltzmann weights of an auxiliary energy-based model. However, ML is generally treated as a `black box' method, since these models are typically of high complexity and abstraction with no physical meaning \cite{1702.08608}. 
The methodologies described in this Letter constitute generative ML because samples from a target distribution are used to train a model that can generalize from those to generate new samples.
Here, though, the auxiliary model is the actual low-energy effective model of interest, and has physical meaning. Importantly, we show that the mapping can be achieved at relatively high temperatures, without having to completely solve the bare Hamiltonian.


\noindent\textit{Partition function condition on effective models.--}
As illustrated in Fig.~\ref{fig:schematic}, the goal is to find an effective model with the same low-energy eigenspectrum, or density of states $\rho(\omega)$, as the bare model. Since the effective model lives in a restricted Hilbert space, its high-energy spectrum is typically more sparse than the bare model. The regime of applicability of the effective model is therefore restricted below some cutoff $E_{\text{cut}}$. At low temperatures $T\ll E_{\text{cut}}$, the thermally-weighted density of states (density matrix spectrum) $q(\omega)=\exp(-\beta \omega) \rho(\omega)$ should therefore agree. This guarantees that the bare and effective models have the same low-temperature thermodynamics, including the same emergent energy scales. Note that if $q_{\rm{bare}}(\omega)\simeq q_{\rm{eff}}(\omega)$ at a given temperature, then the partition functions $\mathcal{Z}=\int d\omega q(\omega)$ necessarily match.

In principle, optimizing an effective model could be achieved by minimizing the difference between the bare and effective probability distributions $P(\omega)=q(\omega)/\mathcal{Z}$ by minimizing their Kullback-Leibler (KL) divergence \cite{kullback1951},
\begin{eqnarray}
\label{eq:KL}
D_{\text{KL}} = \int d\omega~  P_{\rm{eff}}(\omega) \log[P_{\rm{eff}}(\omega)/P_{\rm{bare}}(\omega)] \;.
\end{eqnarray}

ML algorithms based in this way on optimizing with respect to the density matrix are referred to as `quantum Boltzmann machines' \cite{PhysRevX.8.021050}. The problem is that this rigorous prescription only applies in the eigenbasis of the models, and the gradient descent update required to find the optimal effective model involves taking derivatives of Eq.~\ref{eq:KL} with respect to tuning parameters. In most cases this is not practicable, and the ML algorithm itself would need to be run on a quantum computer \cite{Biamonte2017}.

Our central result is that this can be avoided if we restrict our attention to effective models that can \emph{in principle} be derived by a continuous RG transformation from the bare/microscopic model. In particular, the low-energy spectrum of the effective model should remain in one-to-one correspondence with the bare model, with the same quantum numbers. Symmetries of the bare model should be preserved (although the effective model may have larger symmetries). We exclude, for example, a large class of effective models involving a non-interacting quantum gas fine-tuned to trivially reproduce the desired eigenspectrum, or other unphysical models. While an RG-derivable effective model $\hat{H}_{\rm{eff}}=\sum_i \theta_i \hat{h}_i$ may have high physical complexity, its parametric complexity $\{\theta_i\}$ is typically modest. A given effective model has correspondingly modest expressibility in terms of describing different physical systems; the structure of an effective model must be appropriate to the physics being described. This is unlike the standard philosophy for Boltzmann machines that employ an unphysical auxiliary energy-based model to represent $P_{\rm{bare}}(\omega)$, with high expressibility but also high parametric complexity \cite{bengio}.

\begin{figure*}[t]
  \centering
\includegraphics[width=16cm]{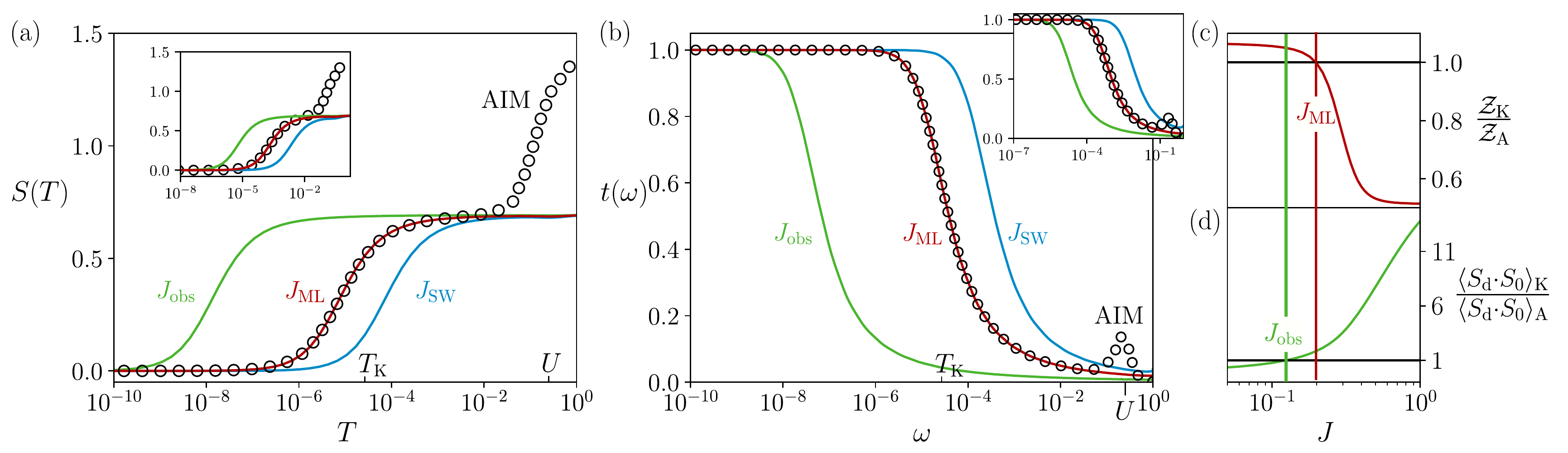}
  \caption{Model ML derivation of the Kondo model from the Anderson impurity model. (a) Impurity contribution to entropy, $S(T)$; (b) $T=0$ spectrum of the t-matrix, $t(\omega)$. AIM results for $U=0.5$ and $8V^2/U=0.25$ shown as circle points. Lines are for Kondo models with $J=J_{\text{ML}}$ optimized by model ML (red lines), $J_{\text{obs}}$ obtained by observable matching (green lines), and $J_{\text{SW}}$ for Schrieffer-Wolff (blue lines). Insets show the same but for reference AIM with $8V^2/U=0.4$. (c) Ratio of partition functions $\mathcal{Z}_{\text{K}}/\mathcal{Z}_{\text{A}}$ for the model ML optimized Kondo model and the reference AIM; (d) corresponding ratio of spin-spin correlators.}
  \label{fig:MML}
\end{figure*}

Since the RG process can be regarded as a `quantum channel' \cite{PhysRevA.92.022330} (a completely positive, trace preserving linear map \cite{PhysRevLett.78.2275}), the partition function is invariant under RG \cite{Cardy1996xt}. An RG-derivable effective model therefore satisfies the condition, $\mathcal{Z}_{\rm{eff}}=\mathcal{Z}_{\rm{bare}}$. Optimization can therefore be done directly on the level of the partition functions.

Our model ML does not perform RG: given a suitable structure for the effective model, the method finds the optimized model parameters by matching partition functions. Even though the partition function is a single number, the method works because we use prior knowledge to restrict the search space. With loss function $L_{\mathcal{Z}}=[\log(\mathcal{Z}_{\rm{eff}})-\log(\mathcal{Z}_{\rm{bare}})]^2$, the gradient descent update for tuning a parameter $\theta_i$ of the effective model is 
\begin{eqnarray}
\label{eq:loss}
\partial L_{\mathcal{Z}}/\partial \theta_i =-2 \beta  [\log(\mathcal{Z}_{\rm{eff}})-\log(\mathcal{Z}_{\rm{bare}})] \times \langle \hat{h}_{i} \rangle_{\rm{eff}} \;.
\end{eqnarray}
The partition functions themselves can be estimated by any suitable method at any temperature $T<E_{\text{cut}}$.


\noindent\textit{Model machine learning for the Anderson model.--}
As a simple but non-trivial proof-of-principle demonstration of  model ML, which can be benchmarked against exact results, we take the Anderson impurity model (AIM) \cite{hewson1993},
\begin{equation}
\label{eq:aim}
\hat{H}_{\text{A}} = \hat{H}_{\text{bath}} 
+ \sum_{\sigma}\epsilon\hat{n}_{d\sigma} + U \hat{n}_{d\uparrow}\hat{n}_{d\downarrow}
+ V\sum_{\sigma} ( d_{\sigma}^{\dagger}c_{0\sigma}^{\phantom{\dagger}}+
c_{0\sigma}^{\dagger}d_{\sigma}^{\phantom{\dagger}} ) 
\end{equation}
where $\hat{H}_{\text{bath}}=\sum_{k,\sigma}\epsilon_k c_{k\sigma}^{\dagger}c_{k\sigma}^{\phantom{\dagger}}$, $\hat{n}_{d\sigma}=d_{\sigma}^{\dagger}d_{\sigma}^{\phantom{\dagger}}$, and $V c_{0\sigma}=\sum_k V_k c_{k\sigma}$. For simplicity we consider particle-hole symmetry $\epsilon=-U/2$, and a flat conduction electron density of states in a band of half-width $D=1$.

The Kondo Hamiltonian is the low-energy effective model  \cite{hewson1993,KWW}, describing impurity-mediated scattering,
\begin{eqnarray}
\label{eq:kondo}
\hat{H}_{\text{K}}=\hat{H}_{\text{bath}} + J \hat{\vec{S}}_d \cdot \hat{\vec{S}}_0 \;, 
\end{eqnarray}
where $\hat{\vec{S}}_d$ is a spin-$\tfrac{1}{2}$ operator for the impurity, and $\hat{\vec{S}}_0=\tfrac{1}{2}\sum_{\sigma,\sigma'}\vec{\boldsymbol{\sigma}}_{\sigma\sigma'}c_{0\sigma}^{\dagger}c_{0\sigma'}^{\phantom{\dagger}}$ is the spin density of conduction electrons at the impurity. For generality, we specify the Kondo conduction electron bandwidth as $D_{\rm{K}}$. For a pure Kondo model, Eq.~\ref{eq:kondo}, the Kondo temperature determining the low-energy physics is given by \cite{RevModPhys.55.331,SM},
\begin{eqnarray}
\label{eq:tk}
T_{\text{K}}(J,D_{\rm{K}}) = \alpha D_{\rm{K}} \sqrt{\rho J} \exp[-1/\rho J + \gamma \rho J] \;,
\end{eqnarray}
where $\rho=1/2D_{\rm{K}}$ is the Fermi level free density of states, $\gamma=\pi^2/4$, and $\alpha=\mathcal{O}(1)$  \cite{SM}. 
The Kondo model is the minimal model, containing only RG-relevant terms consistent with bare symmetries of the AIM \cite{KWW}.  Traditionally, the Kondo model is derived from the AIM by the Schrieffer-Wolff (SW) transformation \cite{PhysRev.149.491}, which perturbatively eliminates excitations out of the singly-occupied impurity manifold. SW yields $J_{\text{SW}}=8V^2/U$ to second order in the impurity-bath hybridization. More sophisticated methods are required to capture non-perturbative renormalization effects neglected by straight SW \cite{haldane1978scaling,*Haldane_1978,tsvelick1983exact,KWW,KEHREIN19961}. The full solution of both Anderson and Kondo models enables comparison of $T_{\rm{K}}$ \cite{tsvelick1983exact,KWW}; the results are often interpreted in terms of a renormalized Kondo bandwidth, $D_{\rm{K}}\rightarrow D_{\rm{eff}} \le D$. SW itself, even to infinite order \cite{SWinf}, does not incorporate bandwidth renormalization. 

Here we use the numerical renormalization group (NRG) method \cite{RevModPhys.47.773,*RevModPhys.80.395,PhysRevLett.95.196801,*PhysRevLett.99.076402,SM} to determine the partition functions of the Anderson and Kondo models ($\mathcal{Z}_{\text{A}}$ and $\mathcal{Z}_{\text{K}}$) at temperature $T$. The Kondo coupling $J$ is optimized by minimizing $L_{\mathcal{Z}}$. NRG results are presented in Fig.~\ref{fig:MML}, comparing the bare AIM (circle points) with effective Kondo models ($D_{\rm{K}}=D=1$): $J$ determined by model ML (red lines), the SW result $J_{\text{SW}}=8V^2/U$ (blue lines), and $J$ obtained by observable matching (green lines, discussed shortly). 
Panel (a) shows the impurity entropy $S(T)$, for $U=0.5$ and $J_{\text{SW}}=0.25$ (inset for $U=0.5$ and $J_{\text{SW}}=0.4$), while panel (b) shows the scattering t-matrix spectrum $t(\omega)$, at $T=0$ for the same parameters. Panels (c,d) show the ML optimization procedure.

Figs.~\ref{fig:MML}(a,b) demonstrate that model ML perfectly determines the true coupling $J$ of the effective Kondo model -- even in the case where an incipient local moment is never fully developed (insets). Deviations between the AIM and Kondo model with $J=J_{\text{ML}}$ set in only at high temperature scales $T\sim U$ (impurity charge fluctuations cannot be described by the Kondo model \cite{hewson1993}).

\begin{figure*}[t]
  \centering
\includegraphics[width=16cm]{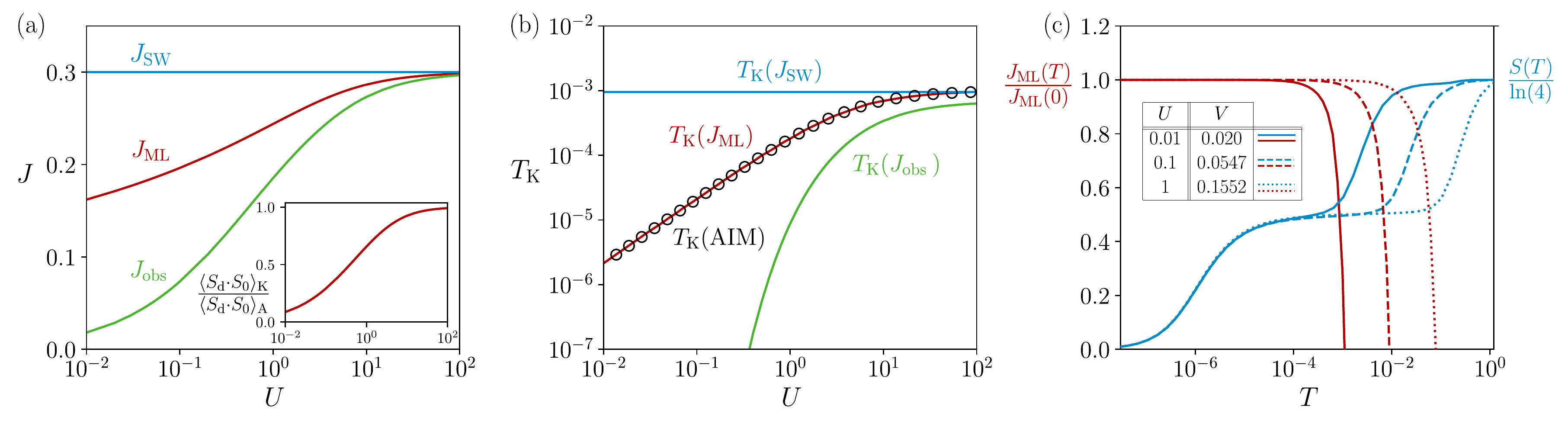}
  \caption{(a) Kondo coupling $J$ optimized by model ML (red line) and observable matching (green line) for a reference AIM with fixed $8V^2/U=0.3$, varying $U$. (b) Corresponding Kondo temperature $T_{\text{K}}$ (circle points for the bare AIM). Inset to (a) shows ratio of spin-spin correlators in the model ML optimized Kondo model and the bare AIM. (c) Temperature-dependence of $J$ obtained by model ML (red lines) for three AIMs with the same $T_{\text{K}}$ but varying $U,V$ [$S(T)$ also plotted for comparison].
  }
  \label{fig:optJ}
\end{figure*}

\begin{figure}[t]
  \centering
\includegraphics[width=8cm]{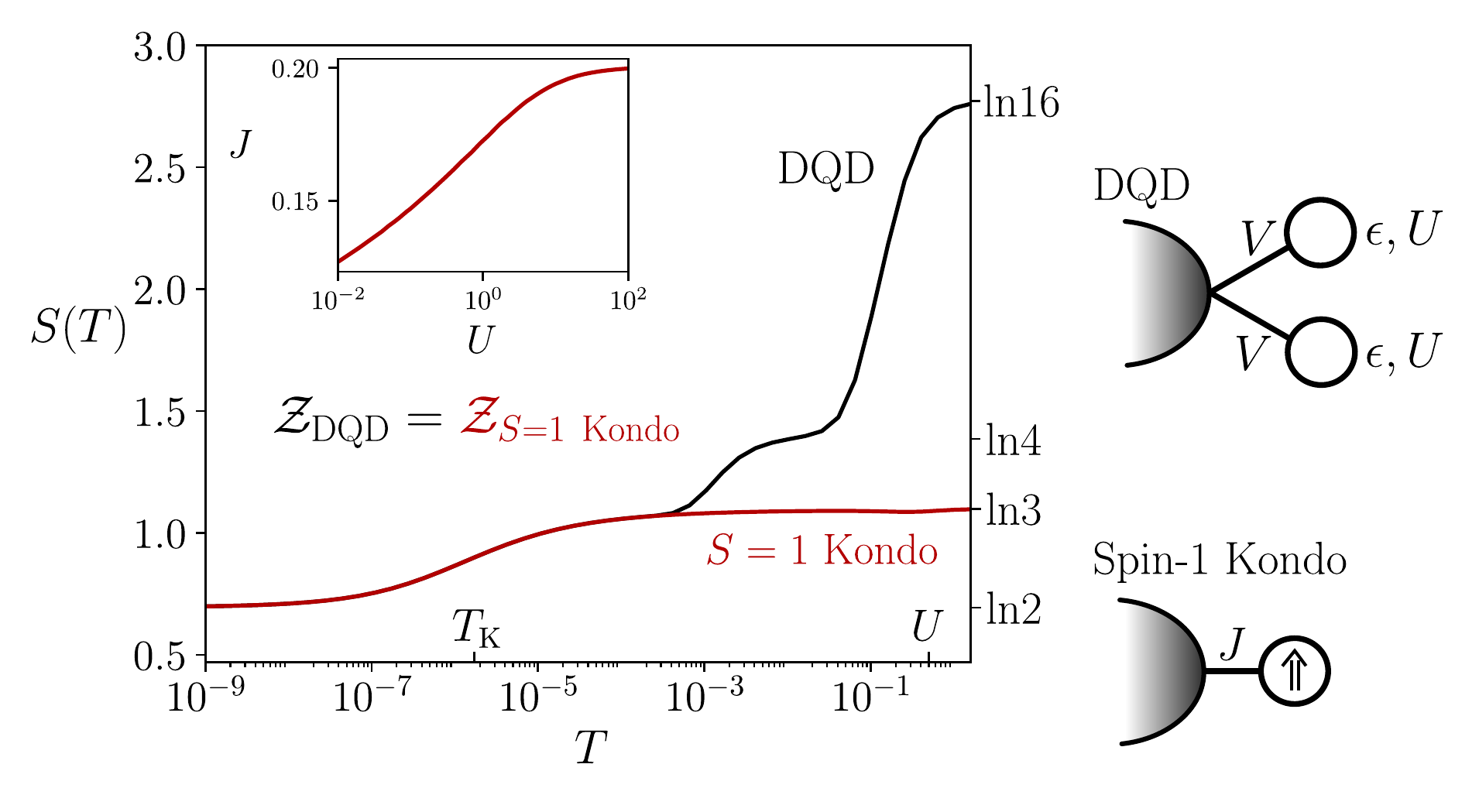}
  \caption{Impurity entropy $S(T)$ vs $T$ for the parallel DQD, with $U=0.5$ and $8V^2/U=0.2$ (black line) compared with an effective spin-1 Kondo model with $J$ determined by model ML (red line). Inset: $J_{\text{ML}}$ vs $U$ for the same $8V^2/U=0.2$.
  }
  \label{fig:dqd}
\end{figure}

The pure SW result substantially over-estimates the coupling, leading to the wrong Kondo scale. For these parameters, our model ML results are consistent \cite{haldane1978scaling,*Haldane_1978} with Haldane's perturbative prediction $D_{\text{eff}} \sim U$ (obtained in the limit $U\ll D$, and neglecting $\rho J_{\text{SW}}$ corrections). However, such analytic estimates are far from straightforward, and not easily generalized. Model ML abstracts and automates the process: Fig.~\ref{fig:optJ} uses model ML to generalize the result beyond the perturbative regime $U\ll D$, while Fig.~\ref{fig:dqd} generalizes to a double quantum dot system. 

In the AIM and Kondo models, the density of states $\rho(\omega)=-\tfrac{1}{\pi}\sum_k \text{Im} G_{kk}(\omega)$, is related to the t-matrix via $G_{kk'}(\omega)=G_{kk'}^0(\omega)+G_{kk}^0(\omega) \mathrm{t}_{kk'}(\omega) G_{k'k'}^0(\omega)$, where $G_{kk'}$ and $G_{kk'}^0$ are the full and free electron Green's functions \cite{hewson1993}. All non-trivial correlations are encoded in the t-matrix spectrum $t(\omega)$ \cite{SM} plotted in Fig.~\ref{fig:MML}(b). Matching the low-energy density of states as per Fig.~\ref{fig:schematic}(b) is therefore equivalent to matching the low-energy t-matrix. However, we have shown that this is achieved automatically by satisfying the simpler condition $\mathcal{Z}_{\text{K}}=\mathcal{Z}_{\text{A}}$.

Fig.~\ref{fig:optJ}(a) shows the evolution of the Kondo coupling $J$ obtained by model ML for a reference AIM with fixed $J_{\text{SW}}=0.3$, but varying $U$ (red line). Panel (b) shows that the Kondo temperature $T_{\text{K}}(J_{\rm{ML}},D)$ of the ML-optimized Kondo model agrees perfectly with the true $T^{\text{AIM}}_{\text{K}}$ of the AIM (circle points). We find \cite{SM}: 
\begin{eqnarray}
\label{eq:MLvsSW}
T^{\text{AIM}}_{\text{K}} &\simeq& T_{\text{K}}(J_{\text{SW}},D)\times \frac{U}{U+3.2D +17.3 J_{\rm{SW}} } \;,\\
&\equiv &T_{\text{K}}(J_{\text{SW}},D_{\rm{eff}}) \nonumber\;,
\end{eqnarray}
where the Kondo bandwidth renormalization is  $D_{\rm{eff}}/D=T^{\text{AIM}}_{\text{K}}/        T_{\text{K}}(J_{\text{SW}},D)$, consistent with known asymptotes $D_{\rm{eff}}\sim U$ for $U$ and $\rho J \ll D$ \cite{haldane1978scaling,*Haldane_1978,tsvelick1983exact,KWW}, and recovering $D_{\rm{eff}}\rightarrow 1$ for $U\gg D$ where pure SW suffices. Inverting Eq.~\ref{eq:MLvsSW} provides an accurate estimate of the true coupling $J$ of the AIM in terms of the pure SW result. 
For more complex systems, a neural network could be used to learn the relationship between bare and effective parameters from sample data of explicit model ML optimizations.

Fig.~\ref{fig:optJ}(c) shows how the results of model ML depend on temperature. We perform optimization of the Kondo coupling $J$ by matching partition functions at temperature $T$ for three reference AIM with the same $T_{\text{K}}$ but different $U, V$. We find that $J_{\text{ML}}$ is robust and essentially constant for all $T\ll U$, where one expects the Kondo model to apply (in practice, $J$ is obtained with less than 3\% error for $U/T>100$). This has the important implication that model ML can be performed using estimates of the partition functions at relatively high temperatures, making it amenable to treatment with e.g.~quantum Monte Carlo methods \cite{BENNETT1976245,VONDERLINDEN199253,PhysRevB.72.035122,RevModPhys.83.349}. Note that for $T\gtrsim U$, the Kondo model is not a good effective model, and the resulting $J_{\text{ML}}$ vanishes as per Eq.~\ref{eq:loss}.


\textit{Double quantum dot (DQD).--}
We now apply model ML to the more complex case of a parallel DQD (Fig.~\ref{fig:dqd}), 
\begin{equation}
\label{eq:dqd}
\hat{H}_{\text{D}} = \hat{H}_{\text{bath}} 
+ \sum_{\alpha,\sigma}\epsilon\hat{n}_{\alpha\sigma} + U \hat{n}_{\alpha\uparrow}\hat{n}_{\alpha\downarrow}
+ V\sum_{\alpha,\sigma} ( d_{\alpha\sigma}^{\dagger}c_{0\sigma}^{\phantom{\dagger}}+
c_{0\sigma}^{\dagger}d_{\alpha\sigma}^{\phantom{\dagger}} ) 
\end{equation}
where $\alpha=1,2$ labels the two dots. The physics of the DQD is much richer than that of the single dot case, due to the interplay between Kondo physics and an emergent RKKY interaction between the dots, leading to an effective underscreened spin-1 Kondo state \cite{PhysRevB.74.045312}.

An effective DQD spin-1 state forms below the emergent scale $J_{\text{RKKY}}$, which is then partially quenched to leave a residual spin-$\tfrac{1}{2}$ and a singular Fermi liquid below $T_{\text{K}}$. As shown in Fig.~\ref{fig:dqd}, model ML finds the correct $J$ to describe the low-temperature physics ($T\ll J_{\text{RKKY}}$). Eqs.~\ref{eq:tk}, \ref{eq:MLvsSW} again hold but with $\gamma=\pi^2/2$ (inset).


\textit{Optimization using observables.--}
ML employing heuristic cost functions based on physical observables might seem appealing if the goal is to reproduce specific observables of the bare model within the simpler description of an effective model. However, this is not always possible in \emph{minimal} effective models. In general, a minimal model optimized to capture the proper low-energy physics cannot reproduce the value of all local observables in the bare model. This is due to information monotonicity along RG flow \cite{PhysRevA.92.022330}.

This result is presaged by the GBF inequality \cite{feynman1998statistical} for the free energy, $F_{\rm{eff}} \le F_{\rm{bare}} + \langle\hat{H}_{\rm{eff}}-\hat{H}_{\rm{bare}}\rangle_{\rm{bare}}$. Differentiating with respect to the coupling constants of the effective model $\hat{H}_{\rm{eff}}=\sum_i \theta_i\hat{h}_i$, we obtain $\langle \hat{h}_{i}\rangle_{\rm{eff}} \le \langle \hat{h}_{i}\rangle_{\rm{bare}}$.
GBF implies that, when optimizing the effective model with respect to $\theta_i$, the corresponding observable $\langle \hat{h}_i \rangle$ is merely \emph{bounded} by its value in the bare model, not necessarily equal to it. ML using observables and ML involving the partition function only agree at the GBF bound.

In the case of mapping AIM to Kondo, we find that the proper effective model ($J$ determined by model ML) yields $\langle \vec{S}_d \cdot \vec{S}_0 \rangle_{\text{K}} \le \langle \vec{S}_d \cdot \vec{S}_0 \rangle_{\text{A}}$, with the GBF bound satisfied only in the SW limit $U\rightarrow \infty$, see inset to Fig.~\ref{fig:optJ}(a).

To compare with model ML, we implement optimization of the effective Kondo model using the observable-based cost function $L_J=[\langle \vec{S}_d \cdot \vec{S}_0 \rangle_{\text{K}}-\langle \vec{S}_d \cdot \vec{S}_0\rangle_{\text{A}}]^2$.~The green lines in Fig.~\ref{fig:MML} show the result of minimizing $L_J$. The Kondo model with $J=J_{\text{obs}}$ has the same impurity-bath spin correlation as the reference AIM, but does not yield the correct low-energy physics or Kondo scale (panels a,b). Panels (c,d) show that $\langle \vec{S}_d \cdot \vec{S}_0 \rangle_{\text{K}}=\langle \vec{S}_d\cdot \vec{S}_0 \rangle_{\text{A}}$ and $\mathcal{Z}_{\text{K}}=\mathcal{Z}_{\text{A}}$ cannot be simultaneously satisfied. Fig.~\ref{fig:optJ}(a) shows how $J_{\text{obs}}$ varies with $U$ for fixed $J_{\text{SW}}$.  Only for $U\rightarrow \infty$ does $J_{\text{obs}}\rightarrow J_{\text{ML}}$. For $U<1$, $J_{\text{obs}}$ is a poor approximation to the true $J$ ($\simeq J_{\text{ML}}$), see Fig.~\ref{fig:optJ}(b).


\textit{Conclusion and applications.--}
We have shown that the parameters of simplified low-energy effective models can be obtained using ML techniques. Optimization on the level of the partition function, estimated at a relatively high temperature, yields the correct low-energy physics for minimal RG-derivable effective models. However, not all local observables are necessarily reproducible in such a model. It remains an open question as to whether \emph{minimally-constrained} effective models, containing higher-order terms beyond the minimal model, are able to capture simultaneously the universal low-energy physics as well as all local observables.

The model ML framework we introduce is general; applications include deriving effective models for complex molecular junctions \cite{Mitchell2017}, and solving inverse problems for rational design. Model ML may also be adapted to find the effective equilibrium problem for non-equilibrium systems \cite{PhysRevB.62.R16271}, or to find simplified/coarse-grained effective descriptions within multi-orbital/cluster dynamical-mean-field-theory \cite{RevModPhys.78.865,*RevModPhys.77.1027,PhysRevB.100.045153}. \\


\begin{acknowledgments}
\emph{Acknowledgments.--} We thank Sudeshna Sen for useful discussions, and acknowledge funding from the Irish Research Council Laureate Awards 2017/2018 through grant IRCLA/2017/169.
\end{acknowledgments}


%


\end{document}



\title{Machine learning effective models for quantum systems: Supplementary Material}

\author{Jonas B. Rigo}
\author{Andrew K. Mitchell}
\affiliation{School of Physics, University College Dublin, Belfield, Dublin 4, Ireland}


\maketitle

\section{Details of NRG calculations}

In this work, we used Wilson's NRG method \cite{RevModPhys.47.773,*RevModPhys.80.395}, formulated in the complete Anders-Schiller basis \cite{PhysRevLett.95.196801}, to calculate the partition functions of Anderson and Kondo models at finite temperature $T$. Numerical results for the impurity contribution to entropy were obtained by standard thermodynamic NRG  \cite{RevModPhys.47.773,*RevModPhys.80.395}, while the dynamical t-matrix and local spin-spin correlator observables were obtained using the full density matrix approach \cite{PhysRevLett.99.076402}. The spectrum of the t-matrix is defined $t(\omega) = -\pi \rho \rm{Im}\sum_{k k'}\rm{t}_{k k'}(\omega)$. All parameters are given in units of the conduction electron bandwidth, here set to $D=1$. Calculations were performed with NRG discretization parameter $\Lambda=2.5$, retaining $N_s=3500$ states at each iteration, and exploiting $Q$ and $S_z$ quantum numbers. Interleaved NRG (iNRG) could be used with model ML in principle \cite{PhysRevB.89.121105, *PhysRevB.93.235101}. We note that for the purposes of model ML, the partition functions need not be calculated exactly: within some given level of approximation applied to both bare and effective models, only the \emph{difference} $\mathcal{Z}_{\rm{bare}}-\mathcal{Z}_{\rm{eff}}$ needs to be estimated to perform the optimization \cite{BENNETT1976245}. Although NRG is used here, the ML methods we describe do not depend on the techniques used to estimate the partition functions or observables.\\


\begin{figure}[h]
  \centering
\includegraphics[width=8cm]{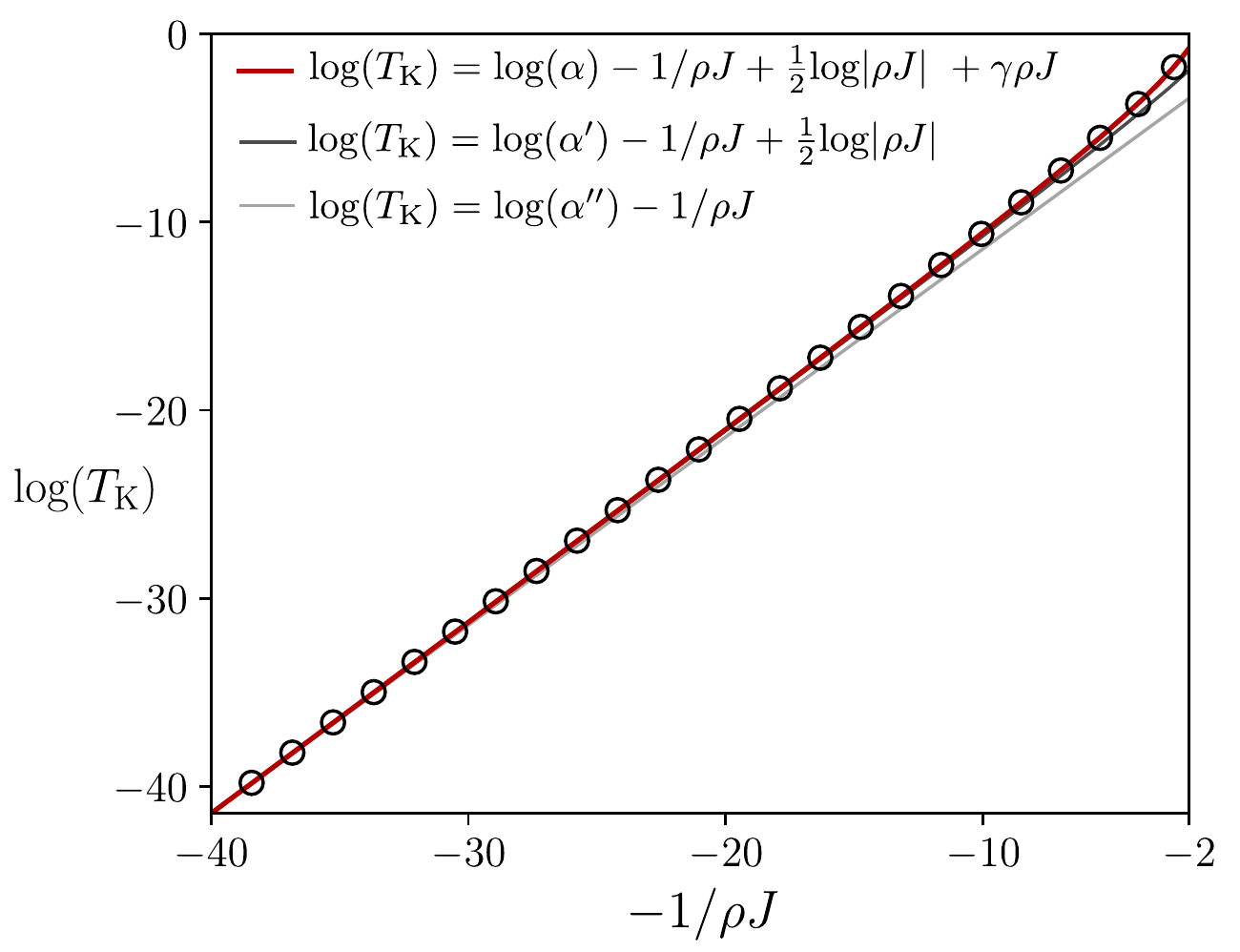}
  \caption{Kondo scale $T_{\rm{K}}$ of the pure Kondo model, Eq.~4, as a function of Kondo coupling $\rho J$, determined from NRG (points), compared with the full analytic result Eq.~5 (red line). $D_{\rm{K}}=1$. The agreement is quantitatively precise over the entire range of $\rho J$ considered. The other lines show the breakdown of lower-order approximations to $T_{\rm{K}}$ at larger $\rho J$. 
  }
  \label{fig:tk_k}
\end{figure}

\begin{figure}[h]
  \centering
\includegraphics[width=9cm]{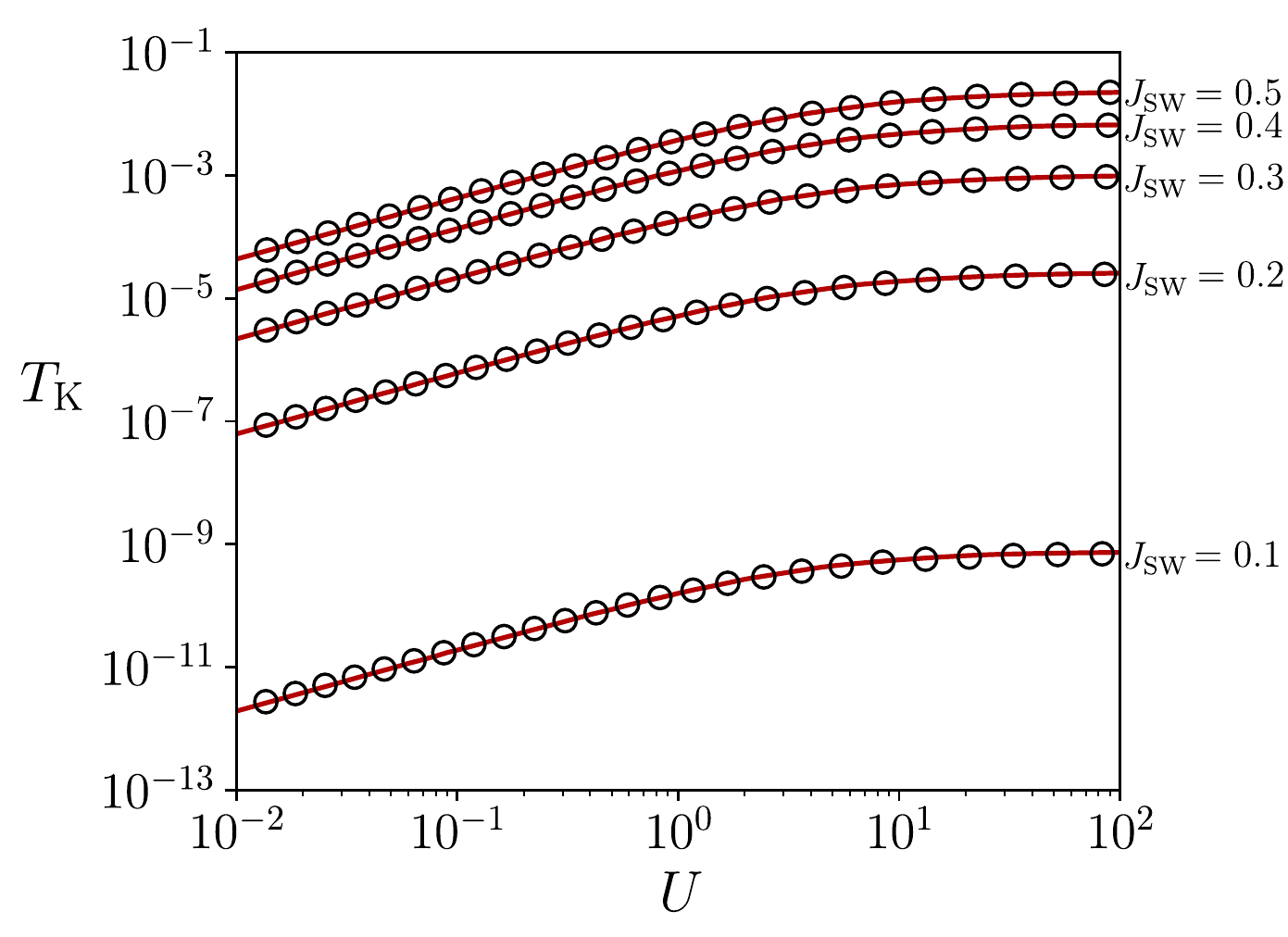}
  \caption{Kondo scale $T_{\rm{K}}$ of the (symmetric) Anderson impurity model, Eq.~3, as a function of interaction strength $U$ for various $J_{\rm{SW}}=8V^2/U$, determined from NRG (points), compared with Eq.~6. $D=1$. Good agreement is seen for all $U$ and $J$ values considered, even in non-perturbative regimes.
  }
  \label{fig:tk_a}
\end{figure}

\section{Kondo scales}
All physical observables are universal functions of $T/T_{\text{K}}$ in the scaling limit of the AIM or Kondo models, and so the definition of the Kondo temperature $T_{\text{K}}$ can be chosen somewhat arbitrarily up to an overall constant. In this work, we choose $\alpha=1.41$ in Eq.~5, so that the conductance half-width-at-half-maximum through an Andersonian dot is precisely at $T=T_{\text{K}}$. As such, this is a natural experimentally-relevant measure. However, for the Anderson / Kondo models, this definition is equivalent to choosing $S_{\text{imp}}(T=T_{\text{K}})\simeq 0.5$, $t(\omega=T_{\text{K}},T=0)\simeq 0.63$, $T_{\text{K}} \chi_{\text{imp}}(T=T_{\text{K}})\simeq 0.1$, or $T_{\rm{K}}=0.236/\chi_{\text{imp}}(T=0)$. In practice we use the latter.

In Fig.~\ref{fig:tk_k} we confirm that the Kondo scale for a pure Kondo model follows Eq.~5 very precisely up to $J/D_{\rm{K}}=0.9$. This is shown using high-quality NRG calculations, in which we retain $N_s=6000$ states per iteration, and perform the $\Lambda\rightarrow 1$ extrapolation \cite{KWW}. Full quantitative agreement requires inclusion of the factor proportional to $\gamma=\pi^2/4$ (note that there is some disagreement/confusion in the literature about this factor, which we definitively pin down here).

In Fig.~\ref{fig:tk_a} we confirm the accuracy of Eq.~6 for the Kondo scale of the (symmetric) Anderson model in terms of $J_{\rm{SW}}$ and $U$, over the whole range $U/D=10^{-2}$ to $10^{2}$, and for $J_{\rm{SW}}/D=0.1, 0.2, 0.3, 0.4, 0.5$.


\bibliography{refs_SM}
